\documentclass[fleqn,usenatbib]{mnras}

\usepackage{newtxtext,newtxmath}

\usepackage[T1]{fontenc}

\DeclareRobustCommand{\VAN}[3]{#2}
\let\VANthebibliography\thebibliography
\def\thebibliography{\DeclareRobustCommand{\VAN}[3]{##3}\VANthebibliography}

\usepackage{graphicx}	
\usepackage{amsmath}	

\newcommand{\fluxunit}{erg\,s$^{-1}$\,cm$^{-2}$}

\title[The JWST FRESCO Survey]{The JWST FRESCO Survey: Legacy NIRCam/Grism Spectroscopy and Imaging in the two GOODS Fields}

\author[P. A. Oesch et al.]{
P. A. Oesch$^{1,2}$\thanks{E-mail: pascal.oesch@unige.ch},
G. Brammer$^{2}$, 
R. P. Naidu$^{3}$\thanks{Hubble Fellow},
R. J. Bouwens$^{4}$,
J. Chisholm$^{5}$,
G. D. Illingworth$^{6}$,
J. Matthee$^{7}$,
\newauthor
E. Nelson$^{8}$,
Y. Qin$^{9,10}$,
N. Reddy$^{11}$,
A. Shapley$^{12}$,
I. Shivaei$^{13}$,
P. van Dokkum$^{14}$,
A. Weibel$^{1}$,
K. Whitaker$^{15,2}$,
\newauthor
S. Wuyts$^{16}$,
A. Covelo-Paz$^{1}$,
R. Endsley$^{5}$,
Y. Fudamoto$^{17,18}$,
E. Giovinazzo$^{1}$,
T. Herard-Demanche$^{4}$,
\newauthor
J. Kerutt$^{19}$,
I. Kramarenko$^{1}$,
I. Labbe$^{20}$,
E. Leonova$^{21,22}$,
J. Lin$^{23}$,
D. Magee$^{6}$,
D. Marchesini$^{23}$,
\newauthor
M. Maseda$^{24}$,
C. Mason$^{2}$,
J. Matharu$^{2}$,
R. A. Meyer$^{25,1}$,
C. Neufeld$^{14}$,
G. Prieto Lyon$^{2}$,
D. Schaerer$^{1}$,
\newauthor 
R. Sharma$^{16}$,
M. Shuntov$^{2}$,
R. Smit$^{26}$,
M. Stefanon$^{27,28}$,
J. S. B. Wyithe$^{9,10}$,
M. Xiao$^{1}$
\\
\emph{\normalsize Affiliations are listed at the end of the paper}
}

\date{Accepted XXX. Received YYY; in original form ZZZ}

\pubyear{2023}

\begin{document}
\label{firstpage}
\pagerange{\pageref{firstpage}--\pageref{lastpage}}
\maketitle

\begin{abstract}
We present the JWST Cycle 1 53.8hr medium program FRESCO, short for ``First Reionization Epoch Spectroscopically Complete Observations". FRESCO covers 62 arcmin$^2$ in each of the two GOODS/CANDELS fields for a total area of 124 arcmin$^2$  exploiting JWST's powerful new grism spectroscopic capabilities at near-infrared wavelengths. By obtaining $\sim$2hr deep NIRCam/grism observations with the F444W filter, FRESCO yields unprecedented spectra at ${\rm R}\sim1600$ covering 3.8 to 5.0 $\mu$m for most galaxies in the NIRCam field-of-view. This setup enables emission line measurements over most of cosmic history, from strong PAH lines at $z\sim0.2-0.5$, to Pa$\alpha$ and Pa$\beta$ at $z\sim1-3$, HeI and [SIII] at $z\sim2.5-4.5$, H$\alpha$ and [NII] at $z\sim5-6.5$, up to [OIII] and H$\beta$ for z$\sim$7-9 galaxies.
FRESCO's grism observations provide total line fluxes for accurately estimating galaxy stellar masses and calibrating slit-loss corrections of NIRSpec/MSA spectra in the same field. Additionally, FRESCO results in a mosaic of F182M, F210M, and F444W imaging in the same fields to a depth of $\sim28.2$ mag (5 $\sigma$ in 0\farcs32 diameter apertures). 
Here, we describe the overall survey design and the key science goals that can be addressed with FRESCO. We also highlight several, early science results, including: spectroscopic redshifts of Lyman break galaxies that were identified almost 20 years ago, the discovery of broad-line active galactic nuclei at $z>4$, and resolved Pa$\alpha$ maps of galaxies at $z\sim1.4$. These results demonstrate the enormous power for serendipitous discovery of NIRCam/grism observations. 
\end{abstract}

\begin{keywords}
surveys -- dark ages, reionization, first stars -- galaxies: formation -- galaxies: evolution -- galaxies: high-redshift 
\end{keywords}



\begin{figure*}
\centering
\includegraphics[width=0.75\linewidth]{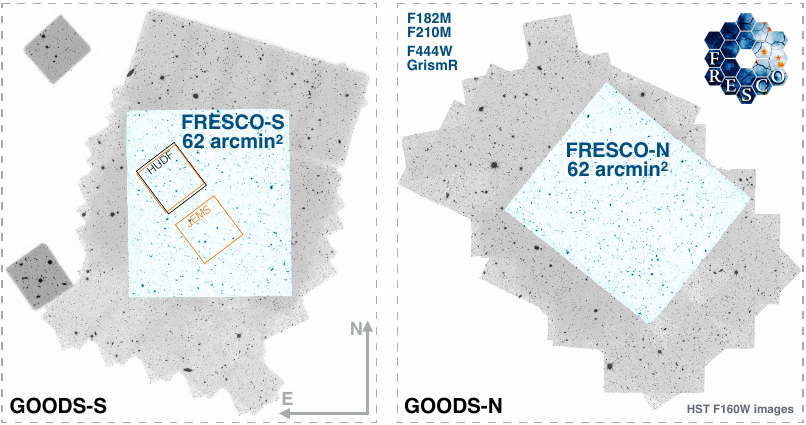}
\caption{ FRESCO was designed to cover the majority of the two CANDELS/Deep regions in the GOODS-S and -N fields, which are among the most valuable extragalactic legacy fields in the full sky. Among others, the FRESCO fields overlap with the medium-band survey JEMS (orange) that covers the original HUDF/XDF (black). The vast amount of ancillary data available in these fields made these fields the obvious choice for spectroscopic follow-up. At the highest redshifts, more than 40\% of all known $z\sim7-8$ candidate galaxies lie in these fields, with GOODS-N featuring an overdensity of $z\sim7-8$ candidates. Additional overdensities are known to exist at lower redshifts, which will be mapped out with these data. The FRESCO mosaics are obtained in a 4$\times$2 mosaic with partial overlap for optimal wavelength coverage. The total area amounts to 7.3$\times$8.5 arcmin$^2$ both with grism spectroscopy and medium band imaging, which further enhances the legacy value of these fields.   }
\label{fig:layout}
\end{figure*}

\section{Introduction}

Revealing the dramatic build-up of galaxies from $z>6$ to the peak of star formation at z$\sim$2-3 is one of astronomy's great achievements with the Hubble and Spitzer Space Telescopes.
Data from these observatories revealed that the first 1 Gyr of cosmic history ($z>6$) was a time of rapid change: soon after the birth of the first stars from metal-free primordial gas, the galaxy population grew rapidly both in star-formation and stellar mass \citep[e.g.,][]{Finkelstein16,Stark16,Oesch18,Bouwens22}.
After $z\sim8$, the star-formation and stellar mass density grew $\sim$10-30$\times$ up to the peak of cosmic star formation at $z\sim2-4$
\citep[see Fig \ref{fig:SFRDLines}; e.g., ][]{madau&dickinson,Bouwens15aLF,Song16,Davidzon17,Furtak21,Stefanon21}.

Even though this general picture of early galaxy build-up is well established, the foundation it stands on remains uncertain. In particular, HST-based analyses only probe the rest-frame ultra-violet (UV) of $z>3$ galaxies, which can lead to a bias against dusty or old galaxies \citep[e.g.,][]{Casey14,Casey18,WangTao19,Xiao22,Barrufet23,Rodighiero23}.
Additionally,  broad-band rest-frame optical imaging can be contaminated by extremely strong emission lines that appear to be common at $z\geq4$, complicating accurate measurements of the underlying continuum and inferences of stellar masses and ages \citep[e.g.,][]{Schaerer09,Labbe13,Stark13,Bisigello19,Endsley22,Stefanon22b}.

The most significant shortcoming before the advent of JWST, however, remained the lack of spectroscopically confirmed redshifts. 
In particular, studies exploring the earliest galaxies in the Epoch of Reionization (EoR) with spectroscopy were sparse before JWST due to disappearing Ly$\alpha$ lines in the neutral IGM \citep[e.g.,][]{Stark10,Treu13,Pentericci14}, and due to low Ly$\alpha$ fractions in general.
Consequently, at $z>7$, only a handful of luminous galaxies had spectroscopic confirmations, based on Ly$\alpha$ or weak rest-frame UV emission \citep[e.g.,][]{Oesch15,Stark17}, or from ALMA ISM lines \citep[e.g.,][]{Inoue16,Hashimoto18,Bakx20,Schouws22,Bouwens22REBELS}. 
This caused extra uncertainty given that the inferences on galaxy abundances at high redshift are statistical in nature. The agreement between candidate catalogs from different teams often remained limited, even when based on the same data: $<$60\% overall, dropping to $\sim30\%$ at the faint end \citep[e.g., compare catalogs from][]{Bouwens15aLF,Finkelstein15}.
Similarly, blind spectroscopic surveys, e.g., with VIMOS or MUSE, up to $z\sim6$ consistently showed that a non-negligible fraction (up to 60\%) of genuine high-redshift galaxies can be missed in traditional Lyman Break color-color selections due to scatter in their observed colors \citep[e.g.,][]{Lefevre15,Inami17}.

With the advent of multiplexed infrared spectroscopy of the James Webb Space Telescope (JWST), all these issues that plagued the exploration of early galaxies can finally be overcome.
Already after just a few months of observations with JWST, extremely early galaxy candidates are now identified out to $z\sim12-16$ \citep[e.g.,][]{Bouwens22,Harikane22,Donnan23,Naidu22a,Finkelstein22a,Atek23,Adams23} and spectroscopically confirmed to $z\sim13$ \citep[e.g.,][]{CurtisLake22,Bunker23,ArrabalHaro23}. Additionally, rest-frame optical photometry and galaxy selections are enabled beyond $z>3$ \citep[e.g.,][]{Barrufet23,Rodighiero23,Nelson23,PerezGonzalez23}.

In this paper, we present the JWST FRESCO survey, short for ``First Reionization Epoch Spectroscopically Complete Observations'' (GO-1895; PI Oesch).  
FRESCO fully exploits the unprecedented grism spectroscopic ability of $JWST$ at $\sim4-5\,\micron$ to obtain a \textit{complete} sample of early star-forming emission line galaxies down to $\sim2\times10^{-18}$\fluxunit\ across cosmic history in two $\sim$62 arcmin$^2$ F444W NIRCam/grism mosaics in the extragalactic legacy fields CANDELS/Deep in GOODS-South and North (Figure \ref{fig:layout}).

FRESCO is designed to obtain a spectrum with $R\sim1600$ for every reionization-era galaxy in the field down to $\sim$0.2-0.5 $L*_{UV}$, probing the strong rest-frame optical emission lines [OIII]+H$\beta$ at $z=6.7-9.0$ and H$\alpha$+[NII] at $z=4.9-6.6$, in addition to other emission lines across the full cosmic history. This includes dust-insensitive Paschen lines at $z\sim1-3$, as well as PAH 3.3$\,\mu$m lines at $z<0.5$. Therefore, the FRESCO observations enable accurate measurements of early star formation and stellar mass build up; provide detailed insights into early, low-metallicity star formation; reveal small scale 3D clustering and measure the contribution of mergers to early galaxy assembly. These key science cases will be discussed in this paper, after first introducing the survey design.

This paper is organized as follows: in Section \ref{sec:design}, we introduce the observational design of the FRESCO grism and imaging survey. In Section \ref{sec:goals}, we describe the main scientific goals of FRESCO, before we end with a summary in Section \ref{sec:summary}. 
Throughout this paper a standard $\Lambda$CDM cosmology is adopted with $\mathrm{H_0=70\,km\,s^{-1}Mpc^{-1}}$, $\Omega_M=0.3$, and $\Omega_{\Lambda}=0.7$.

\begin{figure*}
\centering
\includegraphics[width=0.9\linewidth]{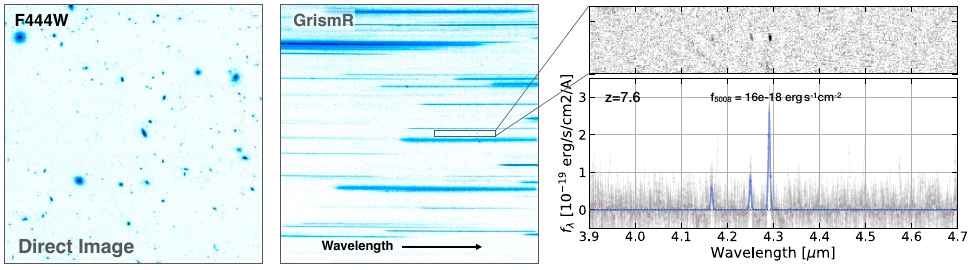}
\caption{Example of FRESCO's spectroscopic strategy. The left panel shows a portion of the F444W direct image in the FRESCO-S field. The middle panel shows the associated GrismR data, from which FRESCO can identify EoR galaxies as emission line sources. Down to UV continuum magnitudes of 26.8, $z=6.7-9.0$ sources are expected to show three emission lines from the [OIII] doublet as well as H$\beta$, resulting in unambiguous redshift identifications. For example, the right panels show the 2D and the extracted 1D spectra of a source in the FRESCO-S field with an unambiguous grism redshift at $z=7.6$. }
\label{fig:grismSim}
\end{figure*}

\begin{table}
    \centering
    \begin{tabular}{|c|c|c|c|} \hline
        Field & RA & DEC & Area [arcmin$^2$]  \\ \hline \hline
        FRESCO-S & 03:32:31 & $-$27:48:04 &  60.9 (grism) \\
        & & & 62.4 (imaging)\\ \hline
        FRESCO-N & 12:36:46 & +62:14:45 & 60.9 (grism)\\
        & & & 62.4 (imaging) \\ \hline \hline
    \end{tabular}
    \caption{Summary of the FRESCO fields and areas.}
    \label{tab:fields}
\end{table}

\section{Observational Design}
\label{sec:design}

The primary goals of FRESCO are to resolve two of the key limitations of our current understanding of the galaxy stellar mass build-up during the epoch of reionization at $z>6$: (1) uncertain redshifts and outlier fractions, and (2)  unknown emission line contribution to rest-frame optical broad-band photometry. In the following, we outline the survey design to overcome these limitations.

\subsection{Slitless Grism Spectra}

The core of FRESCO is NIRCam slitless grism observations, which result in a spectrum for every source in the field-of-view \citep[see, e.g.,][]{Sun22}. FRESCO  follows in the footsteps of numerous successful grism surveys conducted with HST such as 3D-HST \citep{Brammer12,Momcheva16}, GLASS \citep{Treu15}, FIGS \citep{Pirzkal18}, CLEAR \citep{EstradaCarpenter19}, WISP \citep{Atek10}, and GRAPES \citep{Pirzkal04}, but with greater sensitivity, at longer wavelength, and with spectral resolution a factor 10$\times$ higher. 

Unlike traditional slit-based spectroscopy that can only target pre-selected sources, slitless spectra effectively provide a complete sample of emission line selected galaxies. With JWST, the NIRISS instrument provides slitless spectra up to $\sim$2$\mu$m. The NIRCam camera \citep{RiekeNircam,Rieke22} includes a grism mode at longer wavelengths, providing spectra at relatively high resolution, R$\sim$1600, up to 5$\mu$m. In cycle 1, this mode is only used by a few programs, including the GTO/EIGER survey \citep{Kashino22,Matthee22}, as well as the ERS/CEERS program \citep{Finkelstein22a}, who are all using the F356W filter. As we will show below, such grism observations at 3-5\,$\mu$m are extremely powerful probes of galaxy build-up across cosmic history.

\subsubsection{Choice of Filter and Grism}

Grism observations can use the same set of medium and wide filters as NIRCam imaging in its long-wavelength (LW) channel. The choice of filter determines the wavelength range of the spectra. 
The key redshift range that FRESCO targets is $z>6$, i.e., the reionization epoch, where spectroscopic redshifts can be determined from strong rest-optical emission lines such as H$\alpha$+[NII] and [OIII]+H$\beta$. FRESCO therefore obtains 4-5\micron\ grism spectra with the F444W filter. This setup allows us to probe H$\alpha$+[NII] lines for galaxies at $z=4.9-6.6$ and [OIII]+H$\beta$ lines from $z=6.7-9.0$. If bright enough, the [OII] line could in principle even be detected in sources at $z=9.3-12.5$. Hence, the F444W grism spectra result in near complete coverage of the epoch of reionization out to some of the most distant galaxies known in the Universe.

NIRCam is equipped with two different grisms in the LW channel that provide the same spectral resolution, but have a dispersion direction that is rotated by 90 degree relative to each other (GrismR and GrismC). FRESCO is using only one of them: GrismR. The advantages of taking data with both grisms are a better handle on contamination along the dispersion direction and unambiguous identification of sources in the direct image from which the spectra originate. Unfortunately, using GrismC results in a disproportionate increase of overheads, given that out-of-field images need to be taken at 35\arcsec\ offsets, which results in separate visits. As an example, if both GrismC and GrismR were used, FRESCO's survey time would have increased by 15hrs ($\sim$30\%) without any gain in sensitivity. Even with only one grism observation, most shortcomings can be mitigated, as demonstrated by the extremely successful HST grism survey 3D-HST, from which the community has built extensive experience to deal with overlapping spectra \citep[][]{Brammer12}. Due to these excessive overhead costs, FRESCO is obtained with GrismR observations only, similarly to the GTO/EIGER survey (GTO-1243; PI Lilly).

\subsubsection{NIRCam/Grism Mosaic and Wavelength Coverage} 

The wavelength coverage of the NIRCam/grism depends on the location of the source on the detector. At a wavelengths of 3.95\,\micron, the F444W grism and the direct image coincide. However, at other wavelengths, the positions are shifted by 9.85\AA/pixel, varying only slightly across the field. This means that only a certain fraction of sources from the direct image will obtain complete spectral coverage. Additionally, the grism spectra cover sources from outside the field-of-view. 
The GrismR spectra of detector A and B are dispersed in the opposite direction. 
By obtaining a second pointing offset by $\sim1.8$ arcmin, 
FRESCO achieves coverage over 3.9-4.4\,\micron\ over almost the full field of view and partial coverage from 3.9-5\,\micron\ over 4.9 arcmin in x-direction. 
In the y-direction, FRESCO obtained 4 pointings with minimal overlap (thus spanning 8.5\,arcmin). Therefore, each FRESCO field covers 6.7$\times8.5$\,arcmin$^2$ (57 arcmin$^2$) at 3.9-4.4\,\micron\ and  4.9$\times8.5$\,arcmin$^2$ (42 arcmin$^2$) at 3.9-5\,\micron\ (corresponding to 73\% of the full mosaic shown in Figure \ref{fig:layout}). More details on the spectral coverage and the spectral extraction will be provided in Brammer et al., in prep.

\subsubsection{Exposure Times and Sensitivity}

FRESCO is designed to reach a 5$\sigma$ emission-line sensitivity for NIRCam/grism observations of $\sim$2$\times10^{-18}$ \fluxunit\ (for compact sources), which is the expected [OIII]5008 flux of sources at 1 mag below L$_\star$ at $z=7-9$ (based on the LF derived from \citealt{debarros19}; see also \citealt{Matthee22}). 
This is achieved with eight grism exposures of 880s taken with the MEDIUM2 readout mode with 9 groups for each pointing. This results in a total grism exposure time of 7043 s at each position (see also Table \ref{tab:exptimes}).
Four large scale dithers are used, but no sub-pixel dithering. The grism exposures are taken in two sets of these four-point dithers, in each set using a different short-wavelength filter for imaging (see Sect. \ref{sec:imaging}). An example of the resulting FRESCO imaging and spectroscopy in the GOODS-S field is shown in Figure \ref{fig:grismSim}.

\subsubsection{Grism Data Reduction}

Details of the NIRCam/grism data reduction will be provided in a data release paper (Brammer et al, in prep). Briefly, we use the publicly available \texttt{grizli} code\footnote{\url{https://github.com/gbrammer/grizli}}, with slightly modified sensitivity curves and spectral traces based on the v4 grism configuration files provided by Nor Pirzkal\footnote{\url{https://github.com/npirzkal/GRISMCONF}}. \texttt{grizli} is used to reduce all images and spectra and to align them to Gaia-matched reference frames. Given that we are mostly interested in emission line detections, continuum subtracted spectra  are created using a running median filter along each row. Following \citet{Kashino22}, the filter uses a 12 pixel central gap to minimize self-subtraction. For each source in the imaging catalog, optimally extracted spectra are then produced based on the individual input exposures. The spectral catalogs will be released together with  individual science papers and with a future data release paper.

\begin{table}
    \centering
    \begin{tabular}{|l|c|c|c|} \hline
        Filter & Exposures & Time [s] & 5$\sigma$ Depth \\ \hline \hline
 F444W grism & 8 & 7043 & $2\times10^{-18}$ \fluxunit \\ \hline
F444W imaging & 3 & 934 & 28.2 mag \\
F210M imaging & 8 & 3522 & 28.2 mag \\
F182M imaging  & 11 & 4456 & 28.4 mag\\ \hline \hline
    \end{tabular}
    \caption{Summary of exposure times per FRESCO pointing. In total, 2 mosaics of 4$\times$2 pointings were obtained. Line sensitivity estimated for compact sources, integrated over full extension of the line. Imaging depths measured as the 5$\sigma$ rms in 0\farcs32 diameter apertures, without corrections to total fluxes.}
    \label{tab:exptimes}
\end{table}

\subsection{FRESCO's Imaging Legacy}
\label{sec:imaging}
Simultaneous with the grism spectra, FRESCO obtains NIRCam medium band images in two short-wavelength filters (F182M and F210M) to extend the space-based legacy data in these fields to 1.8 and 2.1 $\micron$ down to 28.3 mag (at 5$\sigma$).
These medium-band filters seamlessly extend the NIR wavelength coverage of the HST data in the CANDELS/Deep area, complementing the GTO/JADES NIRCam imaging.
These data improve the UV spectral slope measurement of $z>7$ galaxies and, at lower redshift, allow the community to push rest-frame optical imaging analyses beyond the peak of cosmic star formation, from $z\sim2.5$ to beyond $z\sim4$.

After the grism exposures, three direct and out-of-field images are taken in the F444W filter. These images are needed to associate the spectra with individual galaxies.
The exposure times of these direct images are set to produce a $>4\sigma$ detection of every single galaxy for which a significant emission-line detection is expected. This ensures that spectra can be associated with the imaged galaxies in the field. FRESCO thus obtained 934s exposures in F444W, split over three offset positions, leading to a $5\sigma$-depth of 28.2 mag (see Table \ref{tab:exptimes}).

The medium-band images are taken with the same readout mode as the grism exposures (MEDIUM2, 9 groups), resulting in an exposure time of 3522s each. The F182M filter is used during the three F444W imaging exposures, which are taken with the SHALLOW4 readout pattern with 6 groups. It thus receives an additional integration of 934s. The total exposure times per pointing for each filter are listed in Table \ref{tab:exptimes}.

 \begin{figure}
\centering
\includegraphics[width=0.97\linewidth]{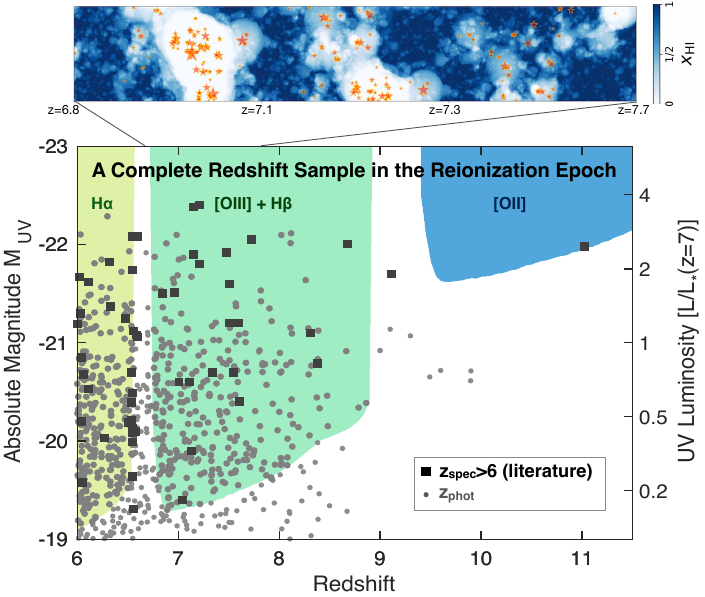}
\caption{The average spectroscopic sensitivity of the FRESCO survey to observe strong rest-frame optical emission lines during the epoch of reionization. Before the advent of JWST, less than 1\% of known sources at $z>6$ had spectroscopically confirmed redshifts. FRESCO is obtaining complete samples with spec-$z$ down to $M_{UV}=-19$ (z$\sim$6) to $-20$ (z$\sim$8) thanks to strong H$\alpha$ and [OIII]+H$\beta$ lines, respectively. This allows us to map out the large-scale structure during cosmic reionization. Simulations predict that ionized regions first appear around overdensities as shown by a slice through the DRAGONS simulation (top panel). The 3D large scale structure during the epoch of reionization will for the first time be revealed thanks to FRESCO redshifts.  }
\label{fig:MuvZspec}
\end{figure}

\subsection{Target Fields} 

FRESCO is split into two separate fields with an equal setup covering two of the most well-studied fields in the sky:  the central regions of the two GOODS fields (North and South; \citealt{Giavalisco96}). These were also covered by the Deep tier of the CANDELS survey \citep{Koekemoer11,Grogin11}. In GOODS-South, the FRESCO field further covers the HUDF/XDF field \citep{Beckwith06,Illingworth13,Ellis13}. Additional imaging over these fields was taken over the years by a very large number of programs. A complete listing of HST programs can be found on the Hubble Legacy Field (HLF) release page\footnote{\url{https://archive.stsci.edu/prepds/hlf/}} (see also \citealt{Whitaker19} and \citealt{Illingworth16}). 

These fields are thus among the most valuable extragalactic legacy fields, especially for distant galaxy science. More than 40\% of all HST-selected $z\sim7-8$ candidate galaxies from blank-field surveys lie in these two areas alone \citep[e.g.][]{Bouwens15aLF,Finkelstein15}. One of the first sources spectroscopically confirmed with Ly$\alpha$ at $z>7$ lies in GOODS-N ($z=7.5$, \citealt{Finkelstein13}) within a spectroscopically confirmed overdensity of $z\sim7.5$ galaxies \citep{Jung20}.
GOODS-N also contains the most distant confirmed galaxy with HST: GN-z11 \citep{Oesch16,Bunker23}, a dusty, red QSO at $z\sim7$ \citep{FujimotoZ7qso}, as well as an enigmatic dusty source, HDF850.1, embedded in another overdensity at $z\sim5.2$ \citep{Walter12}.
By targeting both GOODS fields, FRESCO thus enables a first estimate of the diversity of cosmic structures across cosmis history, well into the EoR.

Both of the FRESCO fields are (partially) observed with the proprietary NIRCam/GTO data from the JADES survey (\citealt{Eisenstein23}; see also \citealt{Helton23}), and with two of the MIRI GTO programs (Rieke et al, in prep.; Ostlin et al, in prep.; see also \citealt{Rieke15,Rinaldi23}).
Additionally, the FRESCO data overlap with the public programs JEMS \citep{Williams23} and NGDEEP \citep{Bagley23}.

\subsection{Data Acquisition}

The FRESCO survey design results in a total science time of 35.5 hrs, for a total of 53.8 hrs including overheads. 
The FRESCO-S data in the GOODS-South field were acquired between Nov 13 and Nov 18, 2022, at an orientation of $\sim$0 deg (V3PA in the range 353.9 to 0.6). 
The FRESCO-N data in the GOODS-North field were obtained between Feb 11 and 13, 2023, at an orientation of 230.5 (V3PA in the range 230.37-230.59). The central position of FRESCO-N was offset slightly from the center of the CANDELS/Deep region in order to capture a few high-profile objects and an expected overdensity of $z\sim7.5$ galaxies, as discussed in the previous section.

 \begin{figure}
\centering
\includegraphics[width=0.97\linewidth]{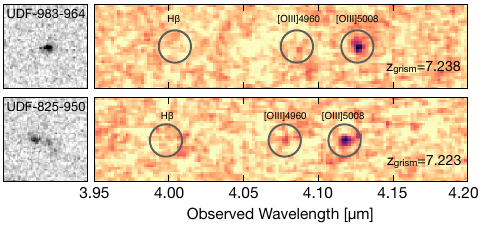}
\caption{An example of how the NIRCam/grism data from the FRESCO survey can finally obtain spectroscopic redshift confirmation for sources that have been identified as high-redshift galaxy candidates almost 20 years prior. The two galaxies shown here have already been detected with the NICMOS camera in \citet{Bouwens04a}. The left stamps correspond to the F444W direct image, while the right panels show the continuum-subtracted grism data. The [OIII] doublet is significantly detected in both cases resulting in an unambiguous redshift confirmation. The H$\beta$ line is found to be weaker for both sources. This example demonstrates the power of NIRCam/grism observations to obtain an emission line selected census of galaxies in the epoch of reionization. }
\label{fig:NICMOSzspec}
\end{figure}

\section{Scientific Goals} 
\label{sec:goals}

FRESCO  exploits the unique capability of the NIRCam/grism to obtain deep spectra at $\sim$4-5$\micron$ for the entire survey field. This enables a wealth of both targeted science, as well as serendipitous discovery.
Below, we discuss some of the most important scientific questions the community can address with these data. 

\subsection{Spectroscopic Census of Early Galaxy Build-up}

\subsubsection{Spectroscopic Redshifts in the Heart of Reionization} 
The first 1 Gyr of cosmic history that constitute the cosmic reionization epoch remain a key unknown in our understanding of the Universe's evolution \citep[see recent reviews][]{DayalReview18,Robertson21}.
Despite enormous efforts, the galaxies in the EoR have eluded almost all attempts at spectroscopic characterization before JWST, apart from a handful of especially bright (or lensed) galaxies. Only two dozen ``normal'' galaxies had confirmed spectroscopic redshifts at $z>7$, out of more than 2000 candidates that were known from prime extragalactic HST legacy fields alone \citep[Fig. \ref{fig:grismSim}; e.g.,][]{bouwens15,Finkelstein15,Atek18}. The main reason for this difficulty was that the primary line for redshift confirmations, Ly$\alpha$, is severely attenuated at $z>6$ due to absorption in the largely neutral intergalactic medium \citep[e.g.,][]{dijkstra14,Mason18}. 
With JWST, this critical shortcoming can finally be overcome through rest-frame optical spectroscopy. FRESCO probes the [OIII]+H$\beta$ emission lines of galaxies at $z=6.7-9.0$ as well as H$\alpha$ for sources at $z=4.9-6.7$. The spectroscopically confirmed redshifts allow us to probe the small scale clustering and 3D correlation function of galaxies during the reionization epoch \citep[e.g.,][]{Endsley20}. From simulations, the progress of reionization is expected to be correlated with the underlying density field (see Fig. \ref{fig:MuvZspec}; \citealt{Qin22}; \citealt{Leonova22}).

Figure \ref{fig:NICMOSzspec} demonstrates the power of FRESCO's NIRCam/grism spectra to measure the redshifts of EoR galaxies. The two highlighted sources are Lyman break galaxies (LBGs) that were identified almost 20 years ago by \citet{Bouwens04a} from the HST/NICMOS images over the Hubble Ultra Deep Field \citep{Thompson05}. With FRESCO, their high-redshift nature is confirmed through the detection of the [OIII]$\lambda\lambda$4960,5008 doublet at $z_\mathrm{grism}=7.238$ and $z_\mathrm{grism}=7.223$, respectively. Importantly, these objects were not selected prior to the observations: all objects in the field are observed spectroscopically, without the need to pre-select targets. This is the main advantage of grism observations and enables a very wide array of science.

\begin{figure}
    \centering
     \includegraphics[width=0.95\linewidth]{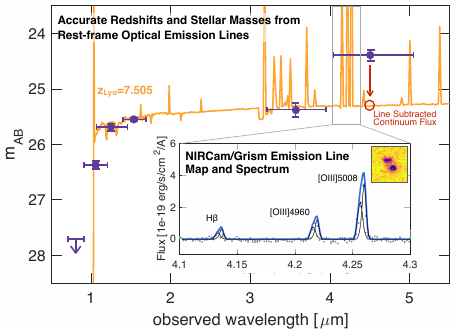}
    \caption{ At $z>5$, the rest-frame optical [OIII]+H$\beta$ lines are found to have observed-frame EWs of $>5000$\,\AA\ -- on average \citep[][]{Labbe10}. This enables extremely efficient spectroscopic confirmation (inset). However, such strong lines contaminate broad-band photometry, as clearly seen in the IRAC CH2 photometry of this source. This can result in stellar mass uncertainties of up to $5-10\times$. FRESCO's spectra enable us to correct JWST's broad-band photometry to obtain true rest-frame optical continuum and hence stellar mass measurements.
 }
    \label{fig:EmlinSubtraction}
\end{figure}

\subsubsection{Accurate Stellar Masses in the Heart of Reionization}

From HST+Spitzer observations, it has been known for several years that the emission line strengths of [OIII] or H$\alpha$ are rapidly increasing toward higher redshift \citep[e.g.][]{Schaerer10,Fumagalli12,Labbe13}, resulting in observed-frame equivalent widths that are $>$5000\,\AA\ on average \citep[e.g.,][]{Smit14,Roberts-Borsani16,debarros19,Endsley21}. This evolution has dramatic consequences for stellar mass estimates that can remain uncertain by factors up to 5-10$\times$ (Fig \ref{fig:EmlinSubtraction}) -- a problem that still applies to NIRCam imaging. \citealt{Bisigello19} estimate \textit{median} correction factors up to 0.87dex, if strong emission lines are not accounted for. FRESCO provides emission line measurements for all sources in the field, enabling precise corrections of the broad-band fluxes for emission line contamination
on a source-by-source basis. 
A prominent example is shown in Figure \ref{fig:EmlinSubtraction}. The source in question still has one of the most distant Ly$\alpha$ detections at $z_\mathrm{Ly\alpha}=7.5$ \citep{Finkelstein13,Jung20}. Thanks to this, it was clear that the excess in the IRAC 4.5$\mu$m band was due to strong rest-frame optical emission lines. Nevertheless, the strength of these lines was uncertain. The FRESCO data shows that the source is actually composed of two components with slightly offset velocities. For both components, the contribution of these emission lines to the broad-band photometry can now be assessed separately.
With these corrections, FRESCO thus provides accurate stellar mass measurements over two key deep fields. Additionally, it will enable the derivation of statistical correction factors for other imaging surveys. %

While these strong rest-optical emission lines can complicate stellar mass estimates, they enable efficient spectroscopic confirmations. 
Therefore, FRESCO's spectra continue to propel the spectroscopic frontier into the heart of cosmic reionization by obtaining redshifts down to 0.2-0.5 $L*$ at $z\sim7-9$ (see Fig \ref{fig:grismSim}). This critically enables measurements of the UV luminosity and stellar mass functions based on pure spectroscopic samples at $z\gtrsim5$.

\begin{figure}
	
	\includegraphics[width=\columnwidth]{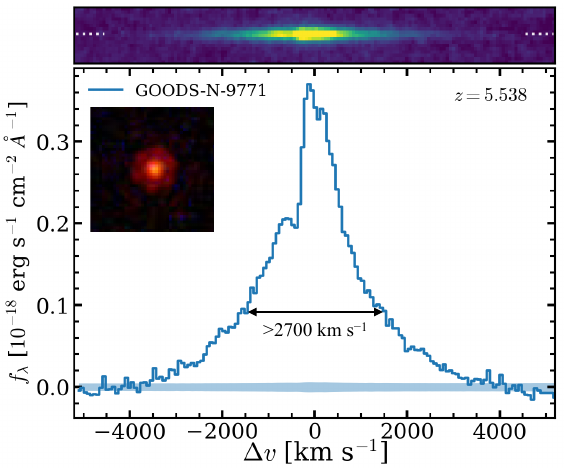}
    \caption{ An example of a serendipitously discovered broad-line, low-mass AGN from FRESCO spectra. The extremely broad H$\alpha$+[NII] lines identify these galaxies as hosting massive black holes (for more information, see \citealt{Matthee23LRD}).  NIRCam/grism observations are uniquely capable of discovering such sources and confirming their number densities in the future. Given the high spectral resolution of $R\sim1600$, such observations also have the power to reveal kinematic information for less extreme galaxies.
}
    \label{fig:LRDexample}
\end{figure}

\subsubsection{Metal-Free Star Formation and the Build-up of Metals at $z>3$} 
A major question for extragalactic surveys with JWST remains whether it is possible to detect primordial (PopIII/zero-metallicity) star formation. 
While  candidate PopIII galaxies at $z\sim7$ have been claimed in the literature \citep[e.g.,][]{Sobral15}, none have been confirmed  (see, e.g., \citealt{Bowler17b,Matthee17CR7}). 
FRESCO has a unique discovery potential for extremely low metallicity candidates at $z>6$. The medium-band imaging at 1.8-2 \micron\ will result in improved UV continuum slope measurements at $z\sim7-9$ enabling the identification of especially blue galaxies for which the grism [OIII] and H$\beta$ line ratios provide an initial gas-phase metallicity estimate ([OIII]/H$\beta$ decreases at $<0.2Z_\odot$; e.g., \citealt{Maiolino08,Inoue11,Curti20}). 
FRESCO thus has the capability to provide promising, extremely metal-poor candidates for future NIRSpec follow-up.
At $z\sim4.9-6.6$, FRESCO continues to trace the build-up of metals through [NII]/H$\alpha$ line ratios in order to constrain the mass-metallicity relations and models of early chemical enrichment.
As such, FRESCO data allows us to trace the transition from extremely metal poor star formation at $z>6$ to more enriched conditions that are found at later times.

\begin{figure}
	\includegraphics[width=\columnwidth]{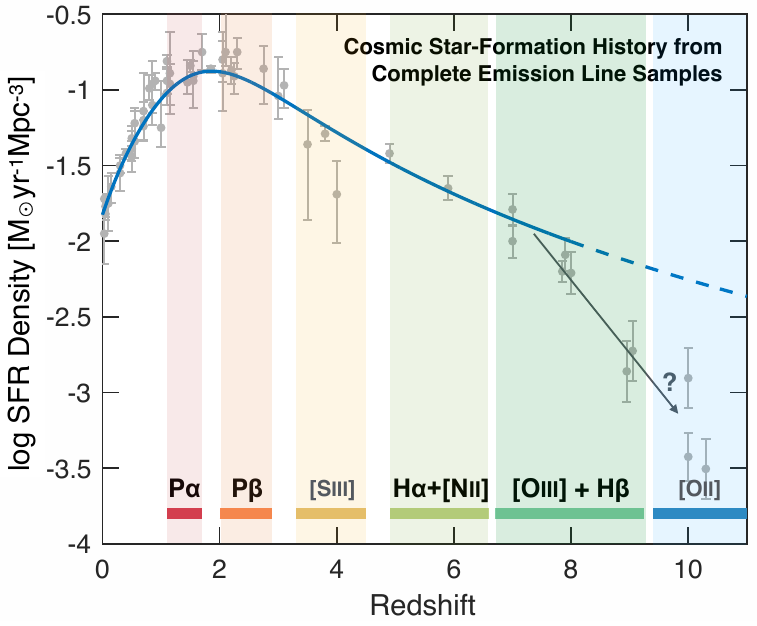}
    \caption{
     FRESCO is designed to return a complete sample of star-forming galaxies based on emission lines across the majority of cosmic history. FRESCO 4-5\micron\ spectra  (a) directly constrain the debated accelerated evolution of the SFRD at $z>8$ based on [OIII]+H$\beta$ samples at $z=$6.9-9.0; (b) follow galaxy build-up through the reionization epoch with H$\alpha$+[NII] measurements at $z=$4.9-6.6; and (c) probe the peak of cosmic SF (z$\sim$1-3) through dust-insensitive Pa$\alpha$ and Pa$\beta$ maps. In the local Universe, FRESCO even covers PAH lines at 3.3 \micron\ for a small number of galaxies (see also Fig \ref{fig:LineRedshift}).
}
    \label{fig:SFRDLines}
\end{figure}

\subsubsection{The Prevalence of AGN among Early Galaxies}

Another central question of extragalactic astronomy is the abundance of active galactic nuclei (AGN) in the early Universe. These sources could provide a non-negligible contribution to reionization \citep[e.g.,][]{Madau15,Giallongo15}, and are expected to explain the abundances of super-massive black holes at later times \citep[e.g.,][]{Volonteri10}. Interestingly, several of the most luminous $z\sim7-9$ galaxies show indications for AGN activity, based on their ground-based rest-UV spectra, suggesting that the AGN fraction could be significant \citep{Laporte17,Sobral18}.
Owing to the R$\sim$1600 spectra, FRESCO can test this scenario directly by enabling the identification of potential AGN through broad emission lines across the full redshift range $z\sim5-9$ based on H$\alpha$, H$\beta$, or [OIII] as well as through indirect methods such as the mass excitation diagram (after a recalibration to $z\sim7-9$ based on $JWST$ spectra; \citealt{Juneau11,Trump13}). 

First JWST spectra have already revealed a significant population of low-mass AGN \citep[e.g.,][]{Kocevski23,Ubler23,Harikane23}, and more candidates are identified through NIRCam imaging \citep[e.g.,][]{Labbe23LRD}. NIRCam/grism observations have the potential to provide a complete sample of such sources. A prominent example from the FRESCO data is shown in Figure \ref{fig:LRDexample}, with an extremely broad H$\alpha$ emission line at $z\sim5$. This source is part of a larger sample of such galaxies identified in NIRCam/grism data \citep{Matthee23LRD}.

\begin{figure}
	\includegraphics[width=\columnwidth]{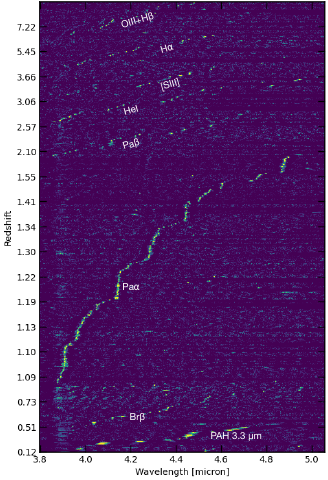}
    \caption{ Line coverage at different redshifts in the NIRCam/grism observations of FRESCO. Shown is a stack of 333 2D spectra in the GOODS-South field for sources with pre-existing spectroscopic redshifts from the literature up to $z\sim6.5$. Beyond that, we show examples of new H$\beta$ and [OIII] lines measured for galaxies at $z>6.8$. Given the large number of spectroscopic redshifts in this field at $z\sim1-2$, Pa$\alpha$ lines currently dominate this figure. Galaxy overdensities are clearly visible. Full spectral extractions of the FRESCO NIRCam/grism data will be described in Brammer et al. (in prep).
}
    \label{fig:LineRedshift}
\end{figure}

\subsection{Star Formation Across Cosmic History}

\subsubsection{Unbiased Star-Formation Rate Indicators:} 
Star formation rate (SFR) is a fundamental observable property of galaxies that is required to trace the growth and formation of galaxies throughout cosmic time. Gold standard SFR indicators are Hydrogen recombination lines, as they emerge from H{\sc ii} regions around the most massive and recently born stars and, hence, trace the almost-instantaneous star formation rate in galaxies \citep{Kennicutt&Evans12}. Aside from Ly$\alpha$, which suffers from uncertainties associated to resonant scattering, Balmer optical lines, such as H$\alpha$ and H$\beta$ are the bright H lines and good tracers of star formation activity. However, these lines are highly affected by dust attenuation, and the uncertainties associated with their attenuation correction factors (e.g., uncertainties in the nebular reddening measurements and attenuation curve assumptions) hamper their potential as accurate SFR diagnostics \citep[e.g.,][among many more]{Fanelli88,Reddy15,Shivaei18}. Pa$\alpha$, on the other hand, is an instantaneous and dust-insensitive SFR indicator, owing to its longer wavelength, making it an important SFR diagnostic, particularly in dusty star-forming galaxies at the peak epoch of cosmic star formation history. FRESCO  takes advantage of the unprecedented near-IR capabilities of JWST to observe Paschen lines (Pa$\alpha$ and Pa$\beta$) and trace optically-thick star formation in large samples of galaxies at $z\sim 1-3$, which is typically missed in optical surveys \citep[see also][]{Finkelstein11,Cleri22,Reddy23}. Another tracer of obscured star formation activity that is accessible to FRESCO is the 3.3\,$\mu$m feature at $z\lesssim 0.5$, which is the emission from PAH dust grains \citep[see, e.g.,][]{Genzel00,Kim12}. 

The FRESCO spectra at $\sim4-5$\micron\ thus have the capability to trace star-formation across the entire cosmic history (see Fig. \ref{fig:SFRDLines}). To illustrate this, Figure \ref{fig:LineRedshift} shows a collage of extracted emission line spectra in the FRESCO-S field. This includes 333 sources that previously had spectroscopic redshifts measured from the literature up to $z\sim6$, in addition to 27 sources for which FRESCO detected bright [OIII]+H$\beta$ lines at $z\sim7$. The spectra are sorted by redshift. Given the availability of previous spectroscopic redshifts the Pa$\alpha$ lines at $z\sim1-2$ dominate the figure. Several overdensities are clearly visible.  The figure shows the power of NIRCam/grism observations to probe star-formation and clustering across cosmic history.

\begin{figure}
\includegraphics[width=0.98\linewidth]{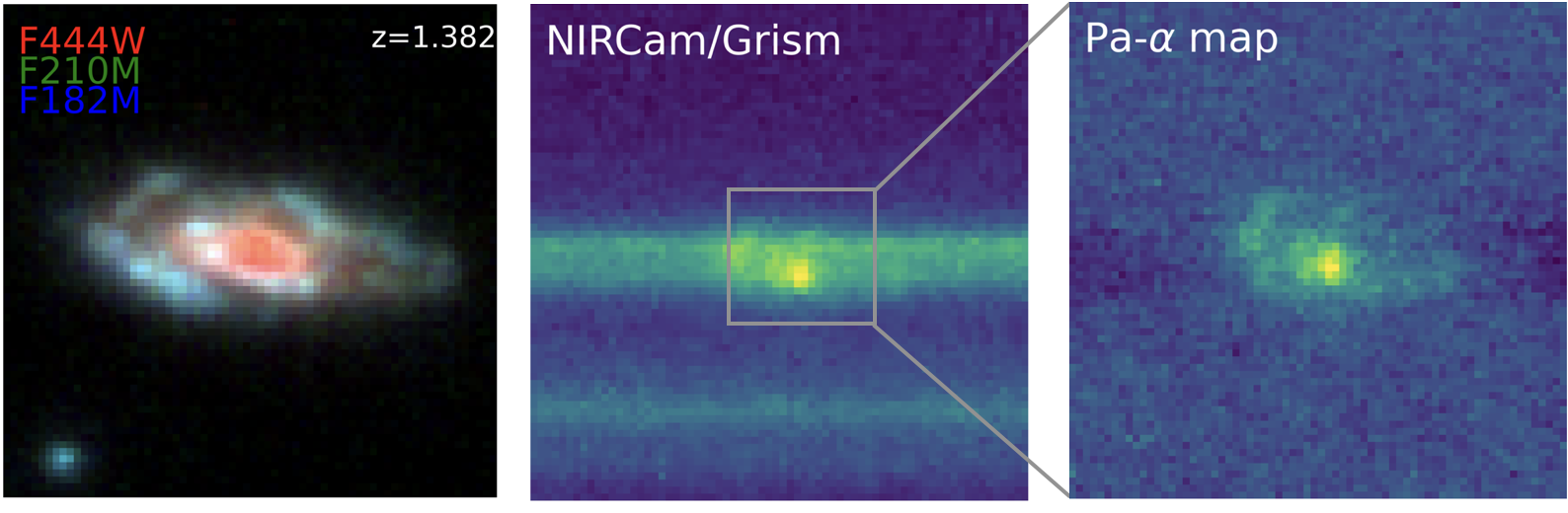}
\caption{ The spatial distribution of SFR from emission line maps across cosmic history as probed by FRESCO. NIRCam/grism data uniquely provide spatially resolved line maps at 4-5\micron.  Shown is the Pa$\alpha$ map from a dusty $z=1.38$ galaxy in the FRESCO-S field. The left panel shows an RGB composite of the FRESCO F182M, F210M, and F444W images. The central panel shows the NIRCam/grism data, from which we extract an emission line map at the Paschen-$\alpha$ wavelength using the \texttt{grizli} tool. The FRESCO dataset thus reveals the sites of star-formation in galaxies.}
\label{fig:PaMap}
  \end{figure}

\subsubsection{Spatial Distribution of Star Formation at Cosmic Noon} 
A unique feature of grism observations is that they result in spatially resolved maps of emission lines and hence star formation. For instance, H$\alpha$ maps from HST grism observations of $z\sim1-2$ galaxies have been used to reveal star-forming clumps and centrally suppressed star formation in massive galaxies, consistent with inside-out growth \citep[e.g.][]{Wuyts13,Nelson16}.
This appears to hold, even when accounting for dust gradients inferred from resolved broadband colors or stacked Balmer decrements \citep[e.g.][]{Tacchella18}. However, ALMA observations of the 870\micron\ dust continuum emission  show a different picture: the dust continuum in many massive galaxies is centrally concentrated \citep{Tadaki20}.  These observations support a different scenario, in which violent dissipative events fuel nuclear star formation.  
Given the uncertainties in H$\alpha$ dust corrections and complications in the interpretation of the dust continuum emission (e.g., due to different heating mechanisms, dust temperature gradients, insensitivity to diffuse emission), an instantaneous and dust-insensitive tracer of star formation is critically needed.
FRESCO provides exactly this, with spatially resolved measurements of the Paschen series lines at $z\sim1-3$.  These measurements allow us to finally answer the question: do massive galaxies at cosmic noon form their last stars in their centers or in their outskirts before they quench?

A first example of a resolved Pa$\alpha$ map in the FRESCO-S field is shown in Figure \ref{fig:PaMap}. The \texttt{grizli}-generated Pa$\alpha$ map indeed reveals a very compact star-forming core for this galaxy that is completely absent in the shorter wavelength imaging. This is embedded in more extended star-formation throughout the galaxy. NIRCam/grism observations such as from FRESCO enable systematic studies of the spatial distribution of star-formation in cosmic noon galaxies.

\begin{figure}
\includegraphics[width=0.95\linewidth]{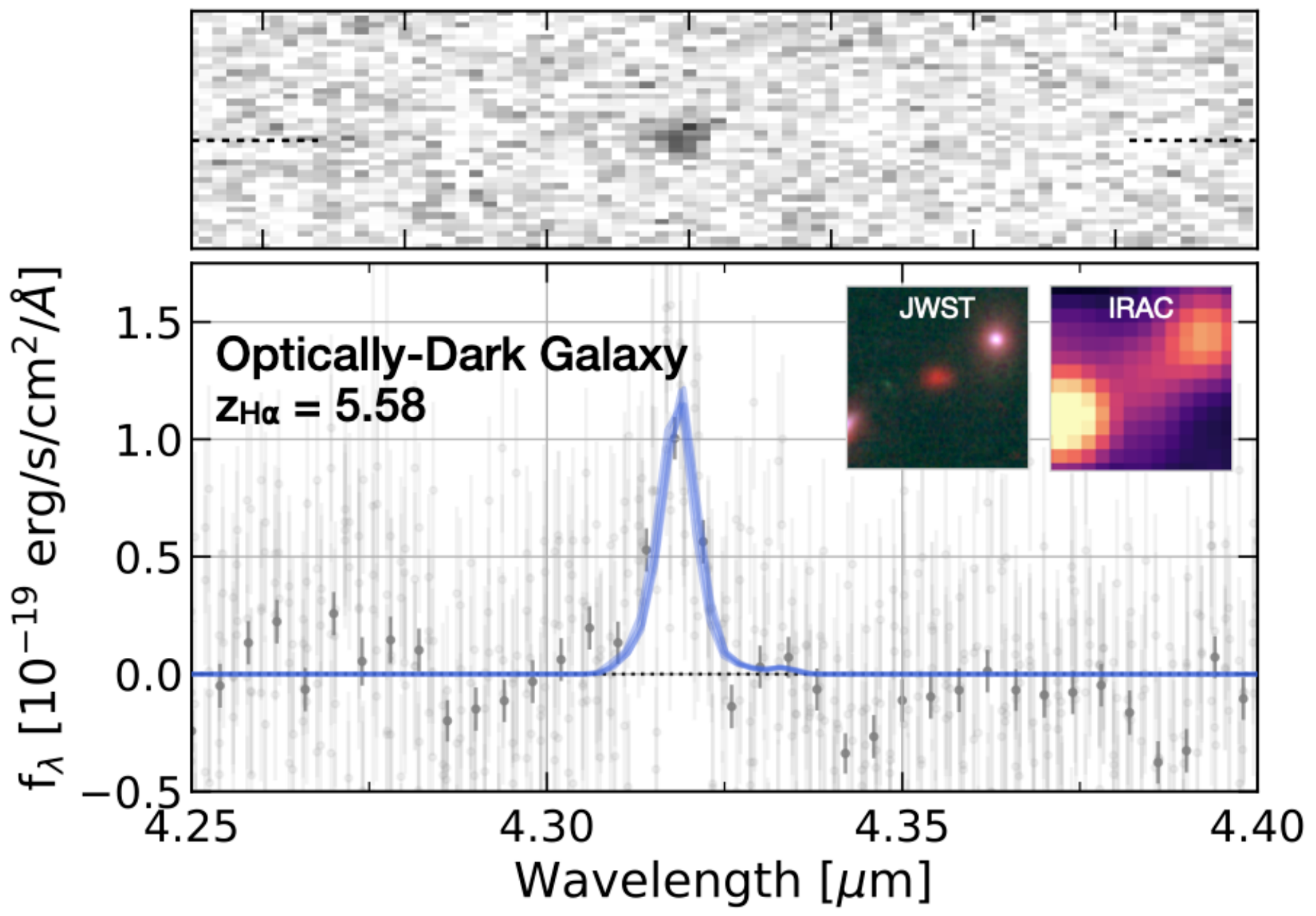}
\caption{FRESCO's imaging and spectroscopy has the power to  reveal the  nature of the enigmatic population of optically-faint galaxies. Shown is an example galaxy from Xiao et al. (in prep.) in the FRESCO-S field. The source was known from ALMA, but its counterpart could not be identified in the low-spatial resolution IRAC imaging. FRESCO's F444W images clearly reveal this galaxy, and with the NIRCam/grism spectra its redshift can be determined through the H$\alpha$ emission line. }
\label{fig:HSTdark}
 \end{figure}

\subsubsection{Illuminating the Complex ISM in Assembling Galaxies}

At $z>5$, galaxies are rapidly assembling, resulting in highly complex interstellar media, as demonstrated by ALMA observations of EoR galaxies, where spatial offsets between the rest-UV continuum, and different ISM lines such as [CII] or [OIII]88\micron\ are common \citep[e.g.,][]{Maiolino15,Carniani17}. FRESCO provides detailed H$\alpha$ (H$\beta$) maps of a large sample of galaxies at $z=4.9-6.6$ ($z=6.7-9.0$) to constrain how and where within galaxies stars are forming during the epoch of reionization \citep[for first examples see the EIGER program][]{Matthee22}. Comparisons between H$\alpha$ and rest-optical continuum sizes provide a first  test of inside-out growth during an epoch when galaxies are still assembling -- rather than fully settled in -- (cold) disks, or may reveal homologous growth with a similar pace of stellar build-up across all radii.

\subsubsection{Illuminating the Dark Side of Star-Formation}

Over the past few years it has become more and more clear that HST-based surveys have been missing a potentially important population of massive, star-forming galaxies at $z>3$ that are sufficiently obscured such that they remained undetected at rest-UV wavelengths 
\citep[e.g.,][]{Caputi15, Franco18, AlcaldePampliega19, WangTao19, Fudamoto21, Xiao22}. These galaxies have been identified either through Spitzer/IRAC or through ALMA observations. However, given the limited photometry available, their photometric redshifts and thus also their nature remained highly uncertain. With JWST we can now finally probe the rest-frame optical wavelengths up to $z\sim9$, which brings this optically-faint galaxy population into view \citep[e.g.,][]{Barrufet23,Rodighiero23,PerezGonzalez23}. FRESCO's imaging and spectroscopy is especially powerful to characterize this enigmatic population of galaxies. The grism spectra enable -- for the first time -- to obtain spectroscopic redshifts for such galaxies thanks to emission line searches. An example of this is shown in Fig \ref{fig:HSTdark} of a galaxy that has previously been detected only at sub-millimeter wavelengths \citep[][Xiao et al. in prep]{Cowie18,Yamaguchi19,GomezGuijarro22,Xiao22}.

\begin{figure}
\includegraphics[width=0.99\linewidth]{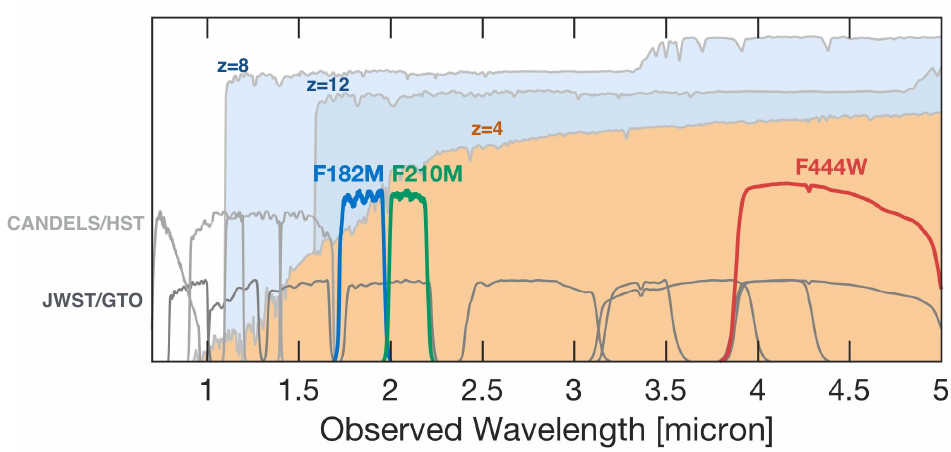}
\caption{FRESCO obtains images in two medium-band filters at $\sim2$\micron\ (and a direct image in F444W). These data seamlessly extend HST's legacy imaging over these fields and complement the GTO data resulting in vast improvements in UV-continuum slope measurements for $z>6$ galaxies and enabling the selection of quiescent galaxies up to $z\sim4$.}
\label{fig:MBimaging}
 \end{figure}

\subsubsection{The Emergence of Quiescent Galaxies}
At $z<2$, the massive galaxy population is dominated by dead, quiescent galaxies, with no significant ongoing star formation \citep[e.g.,][]{Muzzin13,Davidzon17}. However, it is still unknown when such quiescent galaxies first appeared in the Universe. Deep ground-based K-band surveys demonstrated that quiescent galaxies existed since $z\sim4$, however, their exact number densities remain debated \citep{Straatman15,Davidzon17,Forrest20,Valentino20}, mainly because  HST-based surveys are limited to rest-frame optical observations at $z<2.5$. First searches with JWST are already finding and confirming quiescent, low star-formation rate galaxies up to $z\sim5-7$ \citep[see, e.g.,][]{Valentino23,Carnall23,Looser23}. A large population of such sources in the very early Universe challenges simulations \citep[e.g.,][]{Wellons15,Merlin19,Hartley23}, such that the timing of when massive galaxies grow, shut off their star formation and turn quiescent is a very powerful constraint on galaxy evolution models.  

The short-wavelength imaging from FRESCO is designed to complement the JADES/GTO broad-band data in these fields and seamlessly extends HST's legacy data to enable accurate selections of quiescent galaxies up to $z\sim4$ from  Balmer break measurements (see Fig. \ref{fig:MBimaging}). In the SW channel, the imaging filters are the same as obtained with the JEMS survey over the HUDF/XDF.

\subsection{Legacy Science}

The CANDELS/Deep  fields in the centers of GOODS-S and -N are two of the most studied areas in the sky with the most comprehensive ancillary data. 
FRESCO builds on these data and further increases the legacy value of these fields beyond that of the GTO teams by enabling a more complete census of emission line galaxies through blind spectroscopy. Among others, FRESCO provides emission samples for future spectroscopic follow-up with NIRSpec.  Thus, FRESCO's zero-proprietary data enable a wealth of legacy science to be performed by the community. First science papers by the community using FRESCO imaging and spectroscopy have already been published as pre-prints \citep[][]{Laporte23,Helton23}.

\section{Summary}
\label{sec:summary}

This paper provided an overview of the medium program FRESCO -- a NIRCam grism spectroscopy and imaging survey of two fields in the GOODS-North and -South fields, respectively. The power of NIRCam/grism observations are clearly demonstrated with this survey: the grism provides $R\sim1600$ spectroscopy of all sources in the field of view in its long-wavelength channel, while imaging is obtained in a short-wavelength filter at the same time. Adopting this strategy with the F444W filter, FRESCO obtained deep spectra covering 3.8-5.0 $\mu$m, reaching average line sensitivities of 2$\times10^{-18}$\fluxunit\ (5$\sigma$). At the same time, deep F182M and F210M images are obtained reaching $\sim28.2$ mag rms (5$\sigma$ in 0\farcs32 diameter apertures), together with F444W direct images reaching similar depth. The FRESCO observations thus significantly enhance the rich ancillary dataset in these key legacy fields, enabling a vast amount of science by the community. The spectral coverage allows one to probe emission line galaxies across almost the full cosmic history. We have highlighted a few science cases, that showcase the enormous power for serendipitous discovery of such NIRCam/grism observations. Future, wider-area grism programs would thus be able to obtain large, complete samples of rare classes of galaxies that will be very difficult to follow-up, e.g., with NIRSpec  spectroscopy through targeted programs. We hope that the community will use the FRESCO dataset for a large number of scientific discoveries as well as as inspiration for many further JWST programs, including NIRCam/grism observations.

\section*{Acknowledgments}

This work is based on observations made with the NASA/ESA/CSA James Webb Space Telescope. The data were obtained from the Mikulski Archive for Space Telescopes at the Space Telescope Science Institute, which is operated by the Association of Universities for Research in Astronomy, Inc., under NASA contract NAS 5-03127 for JWST. These observations are associated with program \# 1895.

We thank Nor Pirzkal and Fengwu Sun for providing NIRCam grism calibration files, and for helpful discussions.

Support for this work was provided by NASA through grant JWST-GO-01895 awarded by the Space Telescope Science Institute, which is operated by the Association of Universities for Research in Astronomy, Inc., under NASA contract NAS 5-26555.

This work has received funding from the Swiss State Secretariat for Education, Research and Innovation (SERI) under contract number MB22.00072, as well as from the Swiss National Science Foundation (SNSF) through project grant 200020\_207349.
The Cosmic Dawn Center (DAWN) is funded by the Danish National Research Foundation under grant No.\ 140. 

YF acknowledges support from NAOJ ALMA Scientific Research Grant number 2020-16B.

RPN acknowledges funding from JWST programs GO-1933 and GO-2279. Support for this work was provided by NASA through the NASA Hubble Fellowship grant HST-HF2-51515.001-A awarded by the Space Telescope Science Institute, which is operated by the Association of Universities for Research in Astronomy, Incorporated, under NASA contract NAS5-26555.

MS acknowledges support from the CIDEGENT/2021/059 grant, from project PID2019-109592GB-I00/AEI/10.13039/501100011033 from the Spanish Ministerio de Ciencia e Innovaci\'on - Agencia Estatal de Investigaci\'on. This study forms part of the Astrophysics and High Energy Physics programme and was supported by MCIN with funding from European Union NextGenerationEU (PRTR-C17.I1) and by Generalitat Valenciana under the project n. ASFAE/2022/025.

RAM acknowledges support from the ERC Advanced Grant 740246 (Cosmic\_Gas).

CM and GPL acknowledge support by the VILLUM FONDEN under grant 37459.

YQ and JSBW acknowledge support from the Australian Research Council Centre of Excellence for All Sky Astrophysics in 3 Dimensions (ASTRO 3D), through project number CE170100013.

Cloud-based data processing and file storage for this work is provided by the AWS Cloud Credits for Research program.

This paper made use of several publicly available software packages. We thank the respective authors for sharing their work: \texttt{IPython} \citep{ipython},
    \texttt{matplotlib} \citep{matplotlib},
    \texttt{numpy} \citep{numpy},
    \texttt{scipy} \citep{scipy},
    \texttt{jupyter} \citep{jupyter},
    \texttt{Astropy}
    \citep{astropy1, astropy2},
    \texttt{grizli}
    (\citealt{grizli,grizli2}),
    \texttt{EAZY} \citep[][]{Brammer08},
    \texttt{SExtractor} \citep[][]{Bertin96}

\section*{Data Availability}

The HST and JWST image mosaics of the FRESCO fields are already released at MAST as a High Level Science Product via \url{https://doi.org/10.17909/gdyc-7g80}. The spectra are still being calibrated and will be made available together with an upcoming data paper (Brammer et al., in prep.). For updates, please check our webpage:
\url{https://jwst-fresco.astro.unige.ch/} or the MAST page \url{https://archive.stsci.edu/hlsp/fresco/}.

\section*{Affiliations}
\noindent
{\it
$^{1}$Department of Astronomy, University of Geneva, Chemin Pegasi 51, 1290 Versoix, Switzerland\\
$^{2}$Cosmic Dawn Center (DAWN), Niels Bohr Institute, University of Copenhagen, Jagtvej 128, K\o benhavn N, DK-2200, Denmark\\
$^{3}$MIT Kavli Institute for Astrophysics and Space Research, 77 Massachusetts Ave., Cambridge, MA 02139, USA\\
$^{4}$Leiden Observatory, Leiden University, NL-2300 RA Leiden, Netherlands\\
$^{5}$Department of Astronomy, The University of Texas at Austin, 2515 Speedway, Stop C1400, Austin, TX 78712-1205, USA\\
$^{6}$Department of Astronomy and Astrophysics, University of California, Santa Cruz, CA 95064, USA\\
$^{7}$Department of Physics, ETH Z{\"u}rich, Wolfgang-Pauli-Strasse 27, Z{\"u}rich, 8093, Switzerland\\
$^{8}$Department for Astrophysical and Planetary Science, University of Colorado, Boulder, CO 80309, USA\\
$^{9}$School of Physics, University of Melbourne, Parkville, VIC 3010, Australia\\
$^{10}$ARC Centre of Excellence for All Sky Astrophysics in 3 Dimensions (ASTRO 3D), Australia\\
$^{11}$Department of Physics and Astronomy, University of California, Riverside, 900 University Avenue, Riverside, CA 92521, USA\\
$^{12}$Department of Physics \& Astronomy, University of California, Los Angeles, 430 Portola Plaza, Los Angeles, CA 90095, USA\\
$^{13}$Steward Observatory, University of Arizona, Tucson, AZ 85721, USA\\
$^{14}$Astronomy Department, Yale University, 52 Hillhouse Ave, New Haven, CT 06511, USA\\
$^{15}$Department of Astronomy, University of Massachusetts, Amherst, MA 01003, USA\\
$^{16}$Department of Physics, University of Bath, Claverton Down, Bath, BA2 7AY, UK\\
$^{17}$Waseda Research Institute for Science and Engineering, Faculty of Science and Engineering, Waseda University, 3-4-1 Okubo, Shinjuku, Tokyo 169-8555, Japan\\
$^{18}$National Astronomical Observatory of Japan, 2-21-1, Osawa, Mitaka, Tokyo, Japan\\
$^{19}$Kapteyn Astronomical Institute, University of Groningen, P.O. Box 800, 9700 AV Groningen, The Netherlands\\
$^{20}$Centre for Astrophysics and Supercomputing, Swinburne University of Technology, Melbourne, VIC 3122, Australia\\
$^{21}$GRAPPA, Anton Pannekoek Institute for Astronomy and Institute of High-Energy Physics\\
$^{22}$University of Amsterdam, Science Park 904, NL-1098 XH Amsterdam, the Netherlands\\
$^{23}$Department of Physics and Astronomy, Tufts University, 574 Boston Avenue, Medford, MA 02155, USA\\
$^{24}$Department of Astronomy, University of Wisconsin-Madison, 475 N. Charter St., Madison, WI 53706 USA\\
$^{25}$Max Planck Institut f\"ur Astronomie, K\"onigstuhl 17, D-69117, Heidelberg, Germany\\
$^{26}$Astrophysics Research Institute, Liverpool John Moores University, 146 Brownlow Hill, Liverpool L3 5RF, UK\\
$^{27}$Departament d'Astronomia i Astrof\`isica, Universitat de Val\`encia, C. Dr. Moliner 50, E-46100 Burjassot, Val\`encia,  Spain\\
$^{28}$Unidad Asociada CSIC "Grupo de Astrof\'isica Extragal\'actica y Cosmolog\'ia" (Instituto de F\'isica de Cantabria - Universitat de Val\`encia)\\
}



\bibliographystyle{mnras}
\bibliography{MasterBiblio} 





\appendix

\section{Information of Example Galaxies}

In Table \ref{tab:exampleGals}, we provide more details for the sources that are shown in the main text.

\begin{table}
	\centering
	\caption{Source information for the example galaxies shown in Figures \ref{fig:NICMOSzspec}, \ref{fig:EmlinSubtraction},  \ref{fig:LRDexample}, \ref{fig:PaMap},  \ref{fig:HSTdark}}
	\label{tab:exampleGals}
	\begin{tabular}{lccc} 
		\hline
		Fig & RA & Dec & $z_\mathrm{grism}$ \\
		\hline
		\ref{fig:NICMOSzspec}a & 03:32:38.81 & -27:47:07.4 & 7.237$\pm$0.001\\
		\ref{fig:NICMOSzspec}b & 03:32:39.53 & -27:47:17.7 & 7.223$\pm$0.001\\
        \ref{fig:EmlinSubtraction}a & 12:36:37.92 & 62:18:08.7 & 7.507$\pm$0.001\\
        \ref{fig:EmlinSubtraction}b & 12:36:37.87 & 62:18:08.5 & 7.498$\pm$0.001\\
        \ref{fig:LRDexample} & 12:37:07.44 & 62:14:50.3 & 5.538$\pm$0.001\\
        \ref{fig:PaMap} & 03:32:34.04 & -27:50:29.1 & 1.3820$\pm$0.0002\\
        \ref{fig:HSTdark} & 03:32:28.91 & -27:44:31.5 & 5.579$\pm$0.001\\
		\hline
	\end{tabular}
\end{table}



\bsp	
\label{lastpage}
\end{document}